\documentclass[graybox]{svmult}

\usepackage{mathptmx}       
\usepackage{helvet}         
\usepackage{courier}        
\usepackage{type1cm}        
\usepackage{makeidx}         
\usepackage{graphicx}        
\usepackage{multicol}        
\usepackage[bottom]{footmisc}

\usepackage{moreverb}

\makeindex             

\begin{document}
\sloppy

\title*{OMI4papps: Optimisation, Modelling and Implementation for Highly Parallel Applications}
\titlerunning{OMI4papps Status Report} 
\author{Volker  Weinberg, Matthias Brehm, and Iris Christadler}
\institute{Volker Weinberg $\cdot$ Matthias Brehm $\cdot$ Iris Christadler
  \at Leibniz-Rechenzentrum der Bayerischen Akademie der Wissenschaften, \newline
  Boltzmannstr. 1, 85748 Garching bei M\"unchen, Germany \newline
  \email{\{volker.weinberg, matthias.brehm,iris.christadler\}@lrz.de}}
%
%
\maketitle

\abstract*{}
\abstract{This article reports on first results of  
  the KONWIHR-II project \linebreak OMI4papps at the Leibniz Supercomputing Centre (LRZ).
  The first part describes Apex-MAP, a tunable
  synthetic benchmark designed to simulate the performance of typical scientific applications. 
  Apex-MAP mimics common memory access patterns 
  and different computational intensity of scientific codes. 
  An approach for modelling LRZ's application mix is given which makes use of 
  performance counter measurements of real applications running on "HLRB~II", 
  an SGI Altix system based on 9728 Intel Montecito dual-cores.
\newline
  The second part will show how the Apex-MAP benchmark could be used to simulate the performance of two 
  mathematical kernels frequently used in scientific applications: a dense matrix-matrix multiplication
  and a sparse matrix-vector multiplication. The performance of both kernels has been intensively studied on 
  x86 cores and hardware accelerators. We will compare the predicted 
  performance with measured data to validate our Apex-MAP approach.
}

\section{Performance Modelling Using the Apex-MAP Benchmark}
\label{sec:1}

A simple synthetic benchmark with tunable hardware independent parameters that mimics
the behaviour of typical scientific applications is very useful for the evaluation of
new hardware platforms for a certain job mix.
Mapping application performance data measured on a production system 
to specific parameter combinations of the synthetic benchmark 
allows to model the performance of a wide spectrum of applications with a
simple approach. 

To get insight in the performance patterns of the applications running on 
HLRB~II, samples of the most important hardware counters (currently 25
counters) are taken from all processors in 10 minute intervals and are stored
  in a huge database at LRZ. Though the measurements do not only include production runs of optimised user
codes, but also badly optimised programs and test runs etc., the results give
a deep insight into LRZ's job mix and the typical performance of the system. 
Details about the measurement process, the sampling method, the database
scheme and the data analysis can be found in the LRZ technical report 2006-06~\cite{LRZ}.

\subsection{The Apex-MAP Benchmark}

To synthetically model the performance behaviour of LRZ's application mix we
extended the Apex-MAP benchmark (\emph{A}pplication \emph{pe}rformance
\emph{ch}aracterisation project -- \emph{M}emory \emph{A}ccess \emph{P}robe) originally developed by E. Strohmeier \&
H. Shang from the Future Technology Group at the Lawrence Berkeley National
Lab (LBNL), California~\cite{ApexMap1, ApexMap2}. 

The initial idea of the Apex project is the assumption that the performance behaviour of any scientific application can be characterised by a
small set of application-specific and architecture independent performance
factors. Combining these performance factors, synthetic benchmarks that avoid 
any hardware specific model can be designed to simulate typical application
performance. Assuming that the combination of memory accesses and 
computational intensity is the dominant performance factor,
the Apex-MAP benchmark simulates typical memory access patterns of scientific applications.

Concerning the regularity of the memory access, the original Apex-MAP
benchmark focused on random access patterns inside an allocated memory block.
Our implementation also considers strided access patterns, which are common in many scientific applications.
The benchmark written in the style of Apex-MAP has the following 6 parameters:

\begin{description}[Type 1]
\item[M]{The total size of the allocated memory block \texttt{data} in which
  data accesses are simulated, }
\item[L] {the  vector length of data access, 
  (sub-blocks of length $L<M$  starting at \texttt{ind[i]} are accessed in succession), describes the \emph{Spatial Locality}},
\item[$\alpha$]  the shape parameter of  power distribution function ($0\le\alpha\le1$)
  determines the random starting addresses \texttt{ind[i]}, describes the \emph{Temporal
    Locality},
\item[S]{the stride width,}
\item[C]{a parameter used to increase the \emph{Computational Intensity} by calling the
  subroutine \texttt{compute(C)},}
\item[I]{ the length of the index buffer \texttt{ind[]}.}
\end{description}

In the case of strided access only the parameters M, S and C are
relevant. 


The kernel routine for strided access sums up every S-th element of
the allocated memory block \texttt{data[M]}.
\begin{verbatim}
for (int k = 0; k < M/S; k+=1) {
   W0 += c0*data[k*S];
   W0 += compute(C);
}
\end{verbatim}
To increase the computational intensity,
i.e. the ratio of the number of floating point operations and memory accesses,
we added calls to the  subroutine \texttt{compute(C)}:
\begin{verbatim}
double compute(int C){
  double s0,s1,s2,s3,s4,s5,s6,s7;
  s0=s1=s2=s3=s4=s5=s6=s7=0.;
  for(int i=1;i<=C;i++){
    dummy(&s0,&s1,&s2,&s3,&s4,&s5,&s6,&s7);
    s0+=(x[0]*y[0])+(x[0]*y[1])+(x[0]*y[2])+(x[0]*y[3])+
        (x[0]*y[4])+(x[0]*y[5])+(x[0]*y[6])+(x[0]*y[7]);
    s1+=(x[1]*y[0])+(x[1]*y[1])+(x[1]*y[2])+(x[1]*y[3])+
        (x[1]*y[4])+(x[1]*y[5])+(x[1]*y[6])+(x[1]*y[7]);
    ...
    s7+=(x[7]*y[0])+(x[7]*y[1])+(x[7]*y[2])+(x[7]*y[3])+
        (x[7]*y[4])+(x[7]*y[5])+(x[7]*y[6])+(x[7]*y[7]);
  }
  return s0+s1+s2+s3+s4+s5+s6+s7;
}
\end{verbatim}
Performance is usually a mixture of hardware and compiler properties. 
Braces and calls to a \texttt{dummy} routine have been inserted into the \texttt{compute}
routine to assure that the 128 floating point operations in the loop body are
really executed and not cancelled by optimisations of the compiler.
On Itanium the generated assembler code contains 64 consecutive \texttt{fma}
(fused multiply-add)  instructions
which make optimal use of the floating point registers.  
One 128 Byte cacheline is sufficient to hold the two data arrays \texttt{x[8]}
and \texttt{y[8]}. The \texttt{compute}
routine is thus able to run with nearly peak performance on Itanium.

\begin{figure}[b]
\begin{center}
\begin{tabular}{cc}
  \includegraphics[scale=0.3]{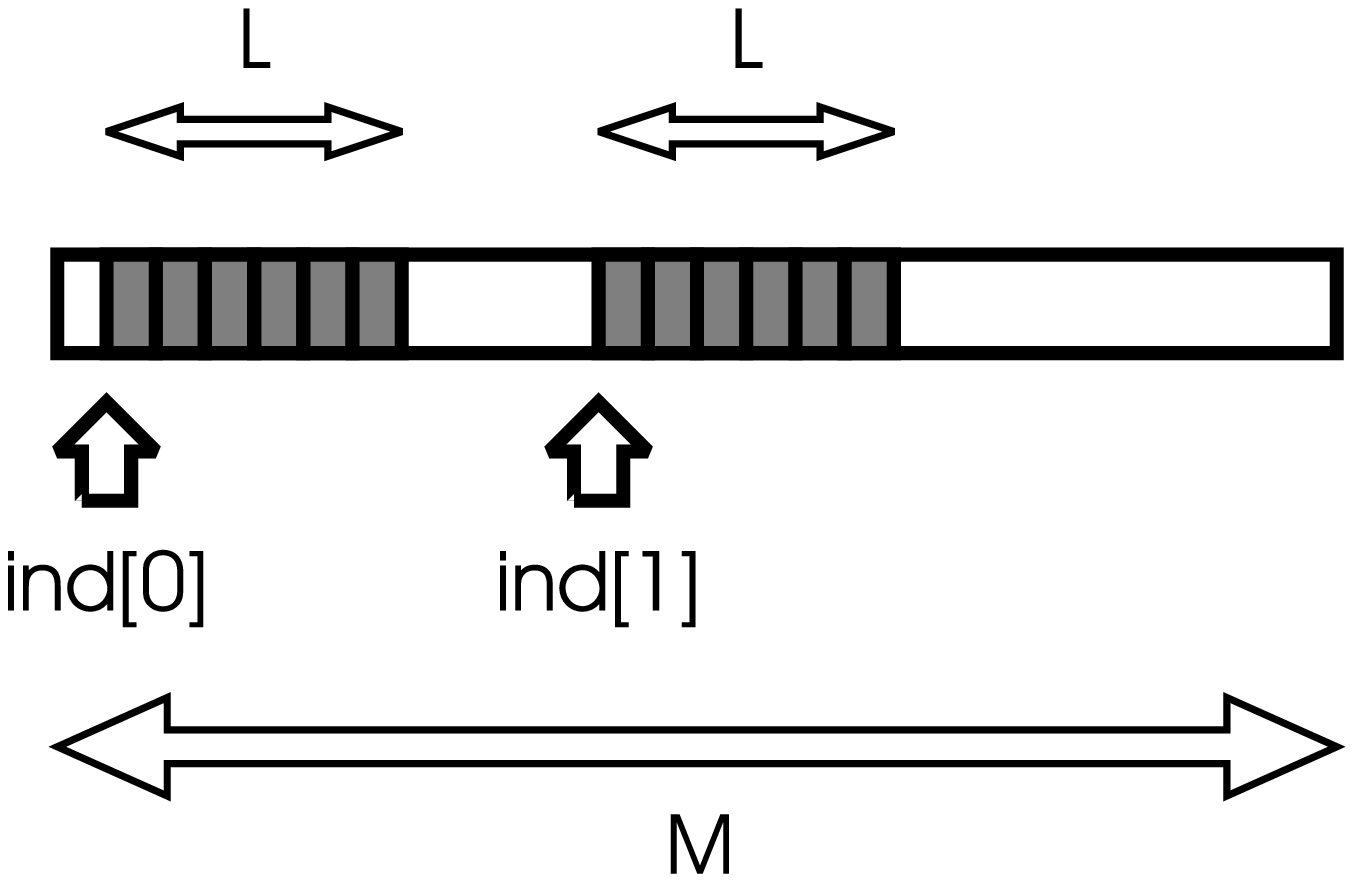}&\includegraphics[scale=0.55]{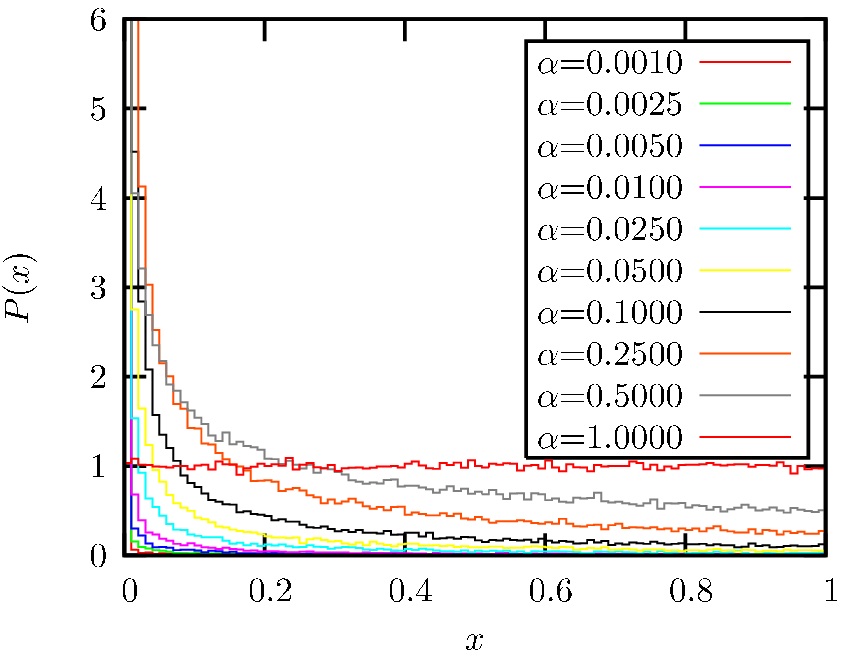}\\
(a) & (b)
\end{tabular}
\end{center}

\caption{Random access pattern of Apex-MAP: the left figure illustrates the
  indexed random access using the index buffer \texttt{ind[]}. The starting addresses
  kept in this array are random numbers drawn from a power distribution
  function with a probability distribution  shown
  on the right for various values of $\alpha$.}
\label{fig:apex}       
\end{figure}

In the case of random access patterns  M, L, $\alpha$,
C and I are the relevant parameters. The kernel routine for random
memory access is:
\begin{verbatim}
for (i = 0; i < I; i++) {
   for (k = 0; k < L; k++) {
     W0 += c0*data[ind[i]+k];
     W0 += compute(C);
   }
}
\end{verbatim}

In this mode I subblocks of length L are accessed. The vector length L is  the
number of contiguous memory locations accessed in succession
starting at \texttt{ind[i]}. L characterises the spatial locality of the
data access. The starting addresses of the subblocks are kept in
the index buffer \texttt{ind[]}. This access pattern is illustrated in
Fig.~\ref{fig:apex} (a).

The starting addresses are random numbers drawn from a power distribution
function and are defined as follows:

\texttt{ind[j] = (L*pow(drand48(), 1/$\alpha$) * (M/L -1)) $\in$ [0;M-L[}

The parameter $\alpha \in$ [0;1] of this distribution function defines the
temporal reuse of data. Figure~\ref{fig:apex} (b) shows the probability
distribution of the power function \texttt{pow(drand48(), 1/$\alpha$)}.
For $\alpha=1$ the random numbers are just deviates with a uniform probability
distribution, while the smaller $\alpha$ is, the more the distribution
function is peaked near 0 and the higher the temporal reuse of data is. For
$\alpha=0$ always the same starting address is used.

\bigskip

\subsection{Comparison of Apex-MAP with Real Application Performance}

To use Apex-MAP for comparing the average memory bandwidth and the floating point performance
of real applications 
the Itanium performance counters \newline
\textit{FP\_OPS\_RETIRED}, \textit{CPU\_OP\_CYCLES\_ALL} and \textit{L3\_MISSES} %
have
been measured and aggregated 
for various combinations of the Apex-MAP input parameters. 
The L3 cacheline size of the Itanium is 128 Bytes and can hold 
16 64-bit (double-precision) values.
The consumed bandwidth between memory
and L3 cache is given by  \textit{L3\_MISSES} $\times$ 128 Bytes. 

\begin{figure}[t!]
\begin{tabular}{cc}
\includegraphics[scale=.3,angle=-90]{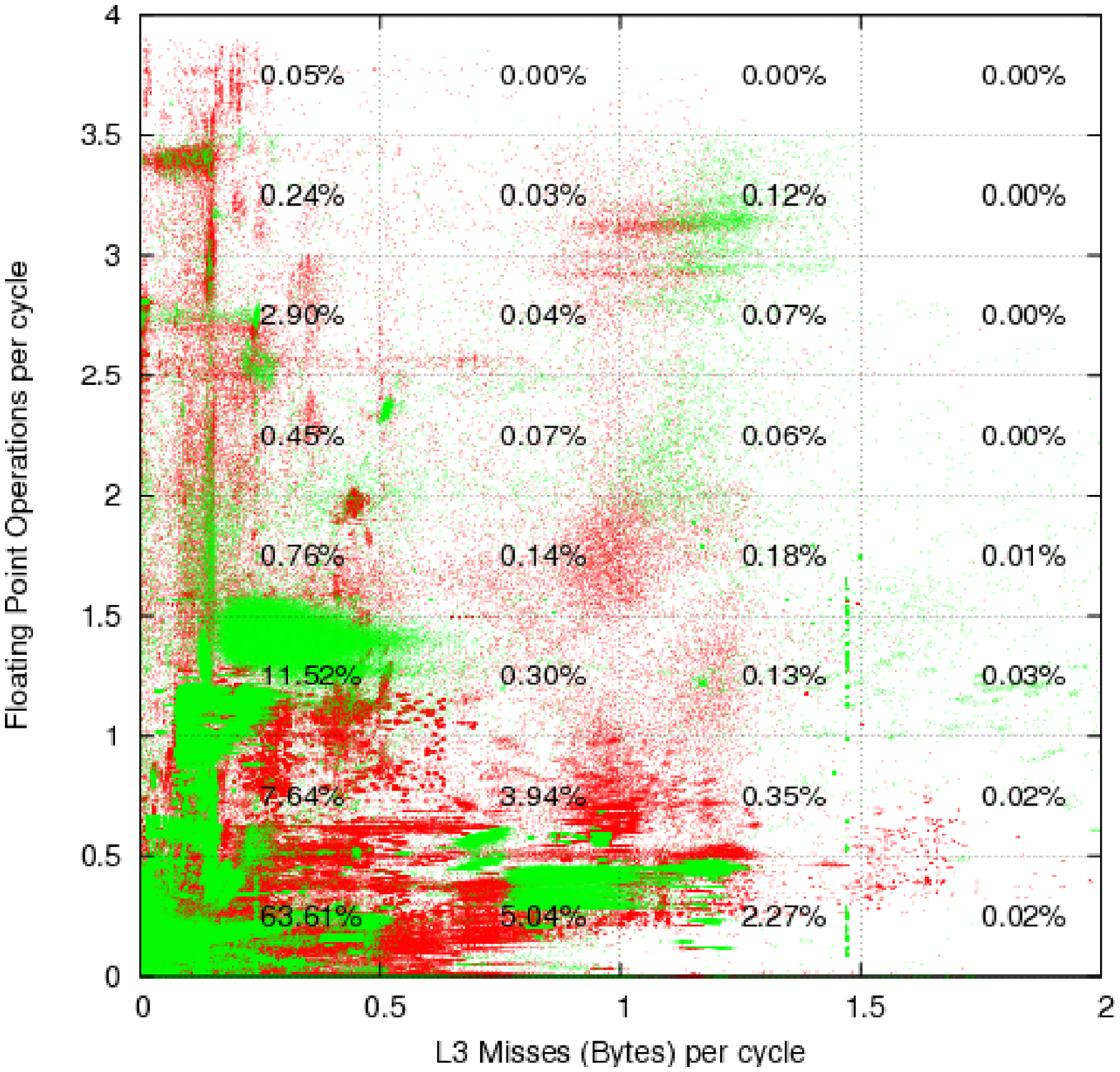}&\includegraphics[scale=.3,angle=-90]{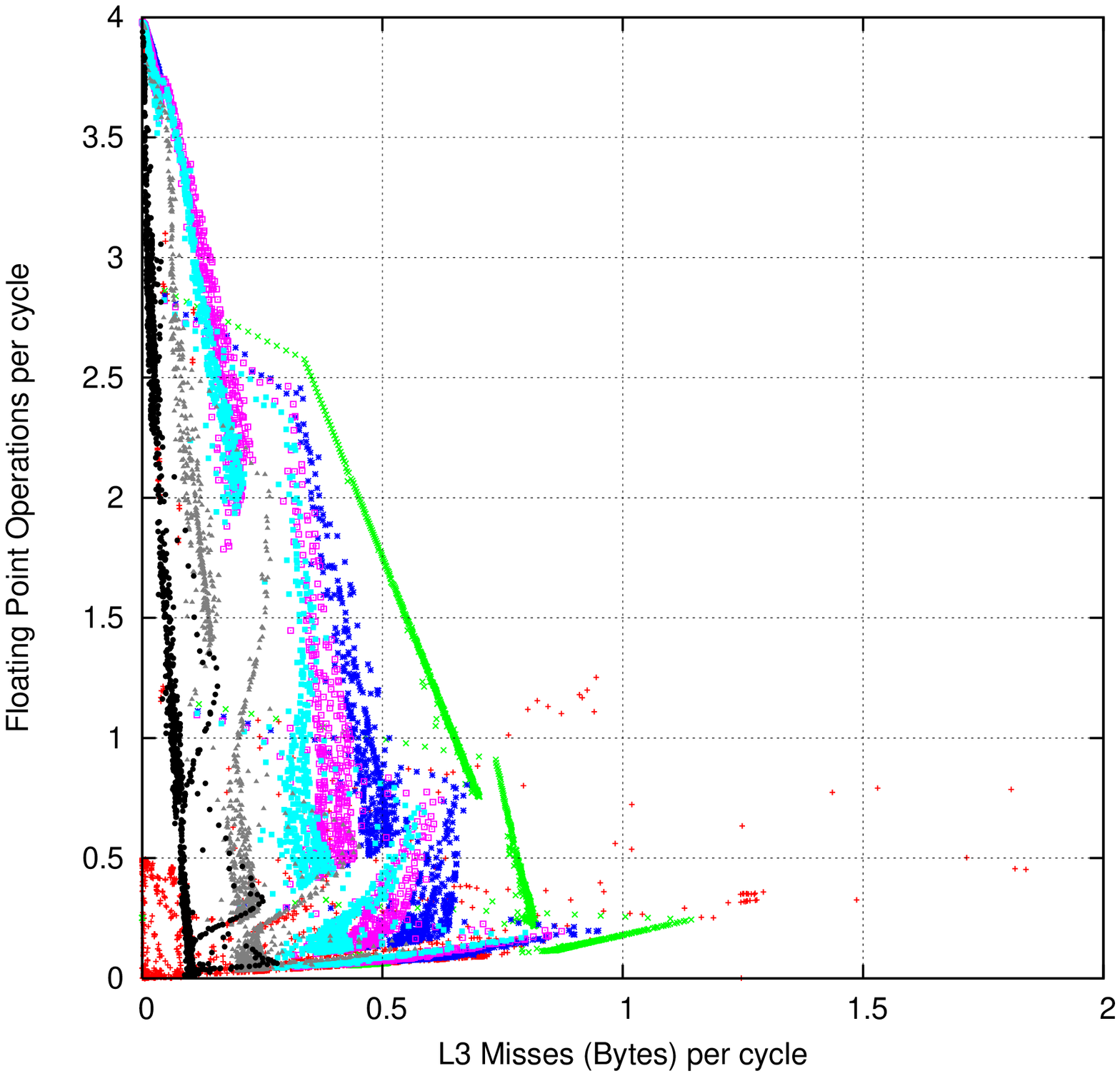} \\
(a) & (b)
\end{tabular}
\caption{Comparison of floating point operations per cycle vs. L3 Misses
  (in Bytes) per cycle for real applications running on HLRB~II (a) 
 and for simulations using the Apex-MAP benchmark with various
  simulation parameters (b). 
  }
\label{fig:counters}     
\end{figure}

Figure~\ref{fig:counters} shows the number of floating point operations per
cycle \linebreak (\textit{FP\_OPS\_RETIRED / CPU\_OP\_CYCLES\_ALL}) versus the
memory bandwidth, expressed by  L3
misses in Bytes/cycle (\textit{L3\_MISSES/CPU\_OP\_CYCLES\_ALL} $\times$ 128
Bytes). Figure~\ref{fig:counters} (a) on the left shows this data for
real applications running on HLRB~II, while Fig.~\ref{fig:counters} (b) on the right
shows data from simulations using the Apex-MAP benchmark with various input parameters.

For the left picture the hardware
  counters were sampled every 10 minutes on all processors of HLRB~II for approximately 3 days
  with a sampling time of 10 seconds. More than 3.2 Mio. samples are taken into account. 
  The average floating point operations per cycle for this 3-day interval is 0.48, which is equal to 
  770 MFlops per core and 12\% of the Itanium's peak performance. 
  The mean for the L3 misses is 0.2 Bytes per cycle. 
 
The parameter space is divided into 32 rectangles of size 0.5 L3 Misses/cycle
$\times$ 0.5 Flops/cycle. The percentage of data points falling into each
rectangle is given. 

For the right picture around 23000 different combinations of the Apex-MAP
benchmark are used. This picture includes various runs, using both random and strided memory access 
as well as serial and parallel runs using OpenMP to cover the same areas as
the measured data on the left. The range of the simulation parameters for Fig.~\ref{fig:counters} (b) is $L<M=1$ GB, 
 $0\le\alpha\le 1$,  $2\le S\le400$, $0\le C\le1000$, I=50.

The two pictures demonstrate that it is possible to model the performance of real applications
by using suitable combinations of input parameters for the Apex-MAP benchmark.
Comparing the two pictures shows that every region in the left
picture with significant percentage (i.e. above 0.5\%) 
 of data points can be covered by a specific
combination of Apex-MAP input parameters. In total, the Apex-MAP runs are able
to cover 98.9\% of the measured real-application performance data.

The measurements  in the left picture are partly based on MPI-parallelised
programs. Performance counters are only implicitly able to measure the
impact of additional communication overheads, e.g. waiting time for external
data on remote processors.  Although Apex-MAP focuses on single processor
performance it is able to mimic the behaviour of parallel applications as long
as the network characteristics stay roughly the same.

\subsection{Modelling LRZ's Application Mix}

It has been shown that Apex-MAP is able to cover the parameter space 
that is attained by real applications. It is assumed that Fig.~\ref{fig:counters} (a)
gives a general overview of the application mix running on HLRB~II. A good indication
for this is given by the fact that the mean MFlops-rate for this 3-day
interval is 770 MFlops per core or 12\% of peak performance, which is a good approximation
of the overall mean application performance of HLRB~II (see also~\cite{HLRBPerfMon}). 
Therefore the weights associated with each rectangle (percentages in
Fig.~\ref{fig:counters} (a))  are used to model the general application mix.

\begin{figure}[h!]
\begin{tabular}{cc}
\includegraphics[scale=.3,angle=-90]{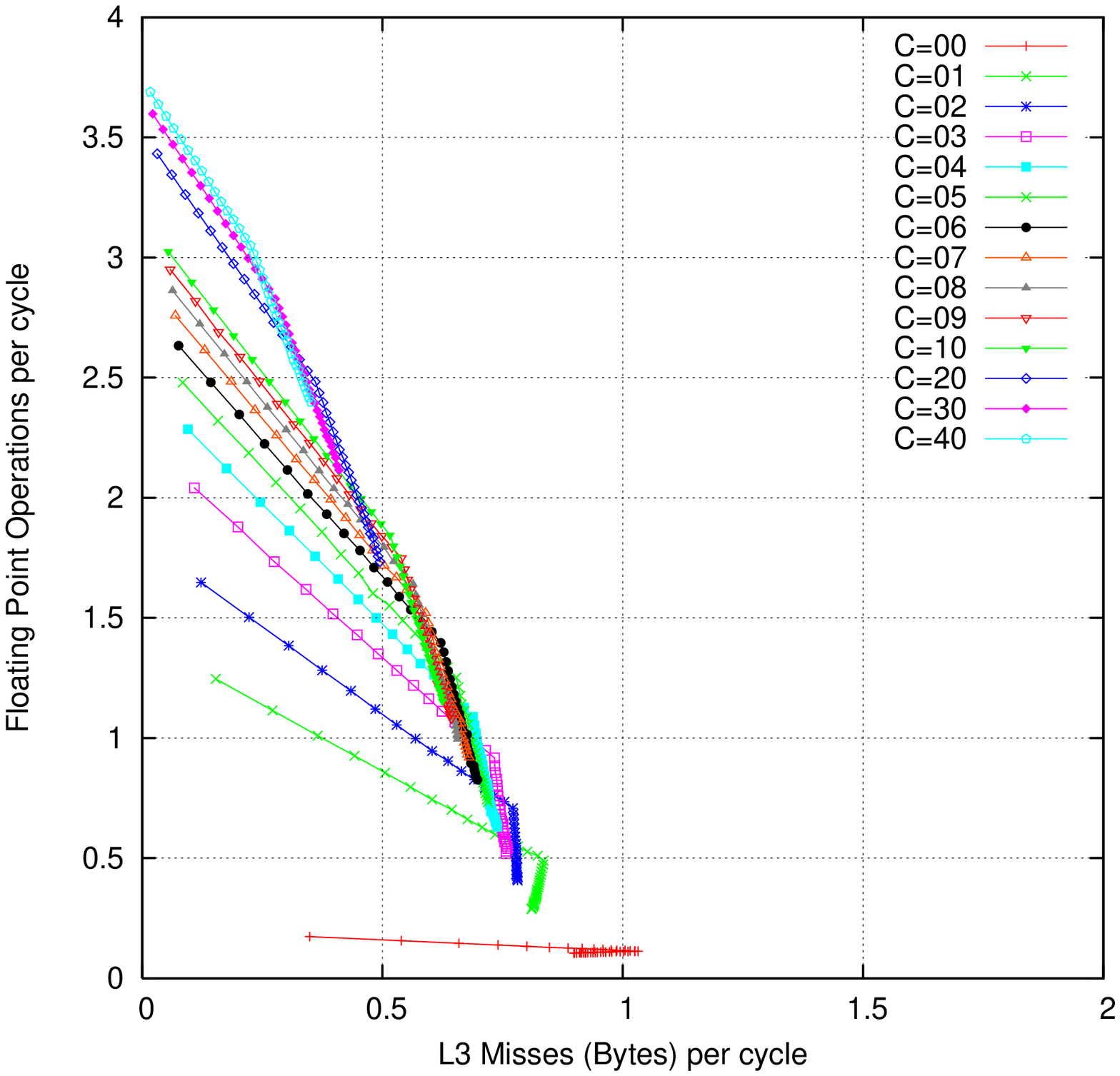}&\includegraphics[scale=.3,angle=-90]{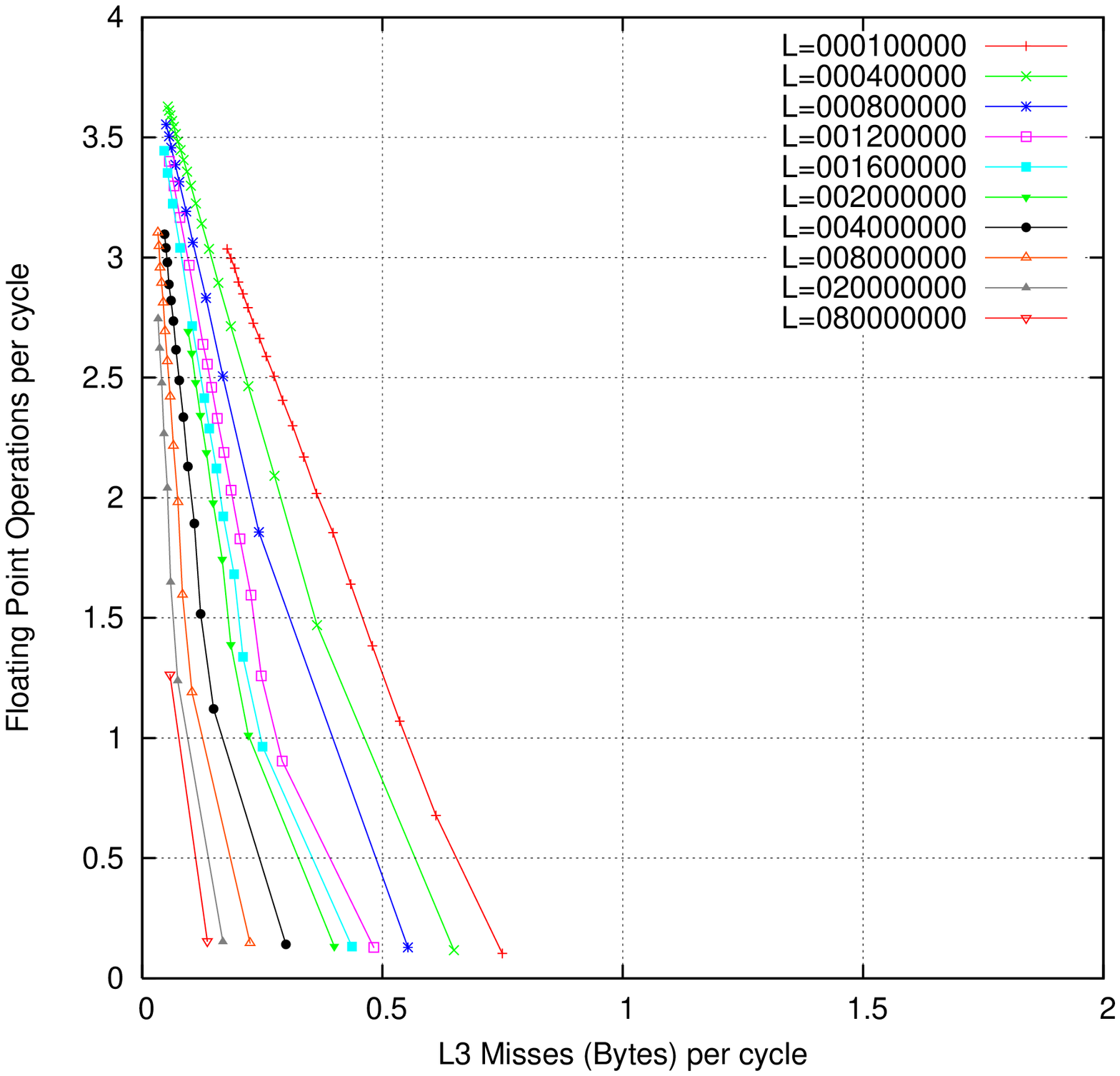} \\
(a) & (b)
\end{tabular}
\caption{Comparison of floating point operations per cycle vs. L3 Misses
  (in Bytes) per cycle for increasing computational intensity C and variations of S for the strided access memory pattern (a) 
  and variations of L for random access memory patterns (b).
  }
\label{fig:stridedVsrandom}    
\end{figure}

Besides the weights for each rectangle, the most suitable combinations of input parameters for Apex-MAP needs to be found.
Figure~\ref{fig:stridedVsrandom} shows the achievable combinations of Flops versus L3 Misses for
different versions of the strided access (a) and random access (b) memory patterns using a serial version of the code. 
Figure~\ref{fig:stridedVsrandom} (a) visualises the common understanding of the influence of a stride
memory access to performance. Every line corresponds to an increase in computational intensity C. As long as the
computational intensity is low (e.g., C stays small), only 1 Flop per clock cycle is possible (which is 
equal to 25\% of peak performance). As the
computational intensity grows larger, the codes are able to run at maximum speed with nearly 4 Flops in every
cycle. As said before, the L3 cacheline size of the Itanium is 128 Bytes;  
16 doubles fit in one cacheline. Therefore with an increase in stride along each line from 1 (contiguous access)
to 16 (access only one item per cache line), the performance drops and stays at a minimum for figures 
above 16 which always need a new cache line. 

\begin{figure}[h!]
\begin{center}
\includegraphics[scale=.35,angle=-90]{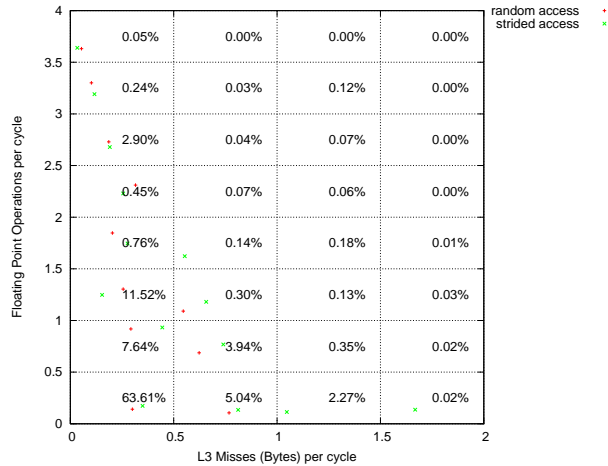} \\
\end{center}
\caption{Chosen data points to model the application mix. These data points represent input parameters
  of the Apex-MAP benchmark. 
  }
\label{fig:inputValues}  
\end{figure}

Figure~\ref{fig:inputValues} shows the data points that have been chosen to model the application
mix of LRZ. The corresponding Apex-MAP input parameters multiplied with the derived weights  
are being used to compute an overall performance of the application
mix in MFlops. Running the adapted Apex-MAP benchmark on HLRB~II yields a performance estimate of
898 MFlops per core. This is quite close to the actual application performance on HLRB~II: 
14\% deviation from the measured 3-day interval.

\section{Validation Using the EuroBen Mathematical Kernels}

The validation of our Apex-MAP version will be done by using two mathematical kernels
typical for many scientific applications.
Within the EC FP7 funded project "Partnership for Advanced Computing in Europe" (PRACE,~\cite{PRACE}), several 
mathematical kernels from the EuroBen benchmark suite~\cite{euroben} 
have been chosen as templates for commonly used scientific applications. To validate Apex-MAP two very
distinct codes have been chosen: 
\begin{itemize}
\item mod2am, a dense matrix-matrix multiplication,
\item mod2as, a sparse CSR (compressed sparse row) matrix-vector multiplication.
\end{itemize}
Within PRACE these codes have been ported to several new languages and architectures; results will be published
on the PRACE website in deliverable D6.6 and D8.3.2~\cite{PRACDel}.
The PRACE surveys analysed the current standards for parallel programming and their
evolution, PGAS languages, the languages introduced as a consequence of the DARPA HPCS
project and the languages, paradigms and environments
for  hardware accelerators.
Performance data has been gathered for various architectures.

The performance of these benchmarks is well known; many different performance runs have been measured,
suitable reference input data sets exists and LRZ was responsible for
the MKL, CUDA and RapidMind ports. 
The first benchmark mod2am has a high computational intensity and is well suited for the use of highly multi-threaded
devices. The second benchmark has a low computational intensity and is a template for codes which will benefit
from a higher memory bandwidth. Using these benchmarks will ensure that Apex-MAP is able to model the two extremes
in terms of computational intensity versus memory access and will allow to 
validate Apex-MAP for the use on hardware accelerators in the future.

\subsection{mod2am: Dense Matrix-Matrix Multiplication}

Several PRACE implementations of the matrix-matrix multiplication are based on the  BLAS Level~3 routine dgemm.
For the x86 implementation the \texttt{cblas\_dgemm} routine from Intel's MKL (Math Kernel Library) has
been used; the CUDA implementation is based on cuBLAS.
The RapidMind implementation uses a code-example from the 
RapidMind developer portal \cite{RMcode} for a  general matrix-matrix multiplication code 
which was slightly adapted. This code is optimised for the use on GPUs. 

\begin{figure}
\begin{center}
\includegraphics[scale=0.7]{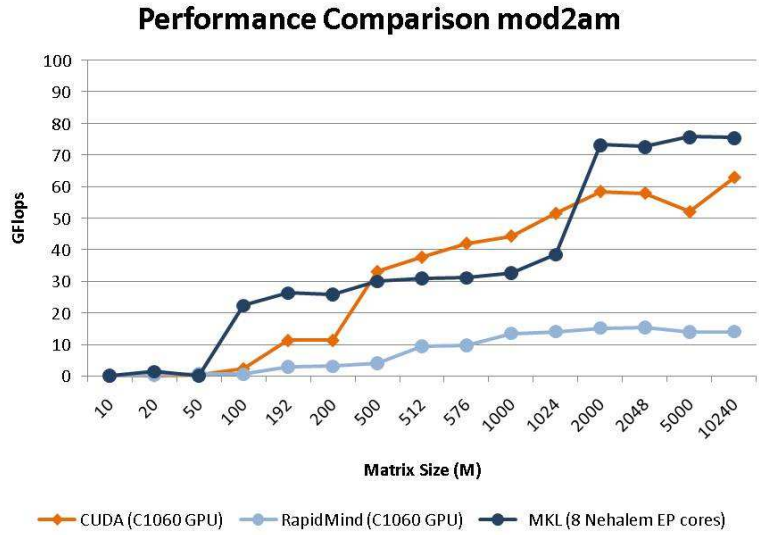}
\end{center}
\caption{Comparison of the performance of the dense matrix-matrix multiplications (mod2am, double-precision) for various
  matrix sizes using RapidMind's CUDA backend, Nvidia's cuBLAS and Intel's Math Kernel Library. 
  The peak performance of one C1060 GPU is comparable to 8 Nehalem EP cores (78 vs. 80~GFlops).
  }
\label{fig:mod2am}
\end{figure}

Figure~\ref{fig:mod2am} shows performance measurements from the PRACE project. It compares
the performance of the CUDA and RapidMind implementations on an Nvidia C1060 GPU, 
which is used in Nvidia's Tesla boxes, with the performance of an MKL version on 8 Intel Nehalem EP cores. 
The reference input data sets are those from PRACE which operate on quadratic matrices. 
They are described in Deliverable D6.6 available
from~\cite{PRACDel} and have been chosen to firstly, represent frequently used problem sizes and secondly, show the dependency
between problem size and performance, especially on hardware accelerators.
The double-precision peak performance of one C1060 (78 GFlops) is comparable to 8 Nehalem cores (80 GFlops).
The diagram shows that the RapidMind implementation is a factor of 4 slower than the highly optimised
cuBLAS library. However, the RapidMind implementation follows roughly the same trend as the CUDA version.

\begin{figure}
\begin{center}
\includegraphics[scale=0.7]{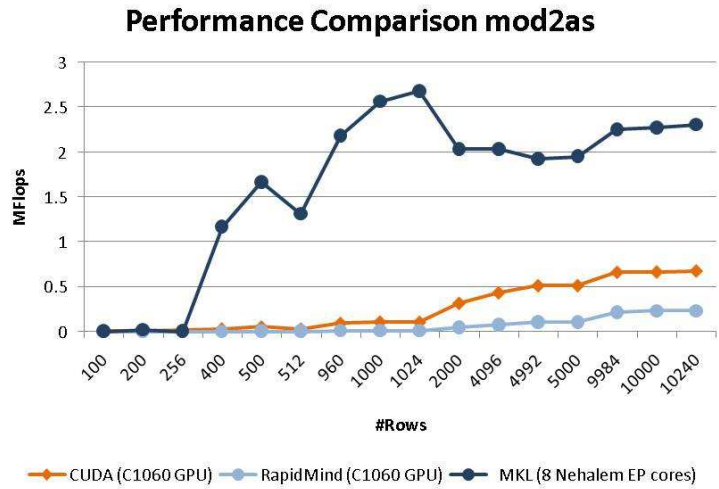}
\end{center}

\caption{Comparison of the performance of the sparse matrix-vector multiplications (mod2as, double-precision) for various
  numbers of rows using RapidMind's CUDA backend, Nvidia CUDA and Intel's MKL.
  The peak performance of one C1060 GPU is comparable to 8 Nehalem EP cores (78 vs. 80~GFlops).
}
\label{fig:mod2as}
\end{figure}

\subsection{mod2as: Sparse Matrix-Vector Multiplication}

In the case of the mod2as benchmark the input matrix  is stored in the
3-array variation of the CSR (compressed sparse row) format. Using this format
only the nonzero elements of the input matrix are stored in one array, and the
other two arrays contain information to compute the row and the column of the
nonzero elements. The entries of the input matrix are computed using a random number generator
but could be reproduced for several runs by using the same seed.

Figure~\ref{fig:mod2as} compares the performance of the RapidMind implementation
with the CUDA and MKL version. The MKL version makes use of a library call
to\newline \texttt{mkl\_dcsrmv}. The CUDA 
implementation is based on the paper "Efficient Sparse Matrix-Vector 
Multiplication on CUDA"~\cite{CUDAmod2as}. A description of the RapidMind implementation
can be found in Deliverable D8.3.2 at~\cite{PRACDel}. 
Again, the reference input data sets from PRACE have been used; all data sets contain quadratic matrices of
different sizes and fill ratios. The diagram shows that the RapidMind version
is a factor of 3 slower than the optimised CUDA version. The trend of both is very similar.

\subsection{Validation of Apex-MAP}

\begin{figure}[h!]
\begin{tabular}{cc}
\includegraphics[scale=.26,angle=-90]{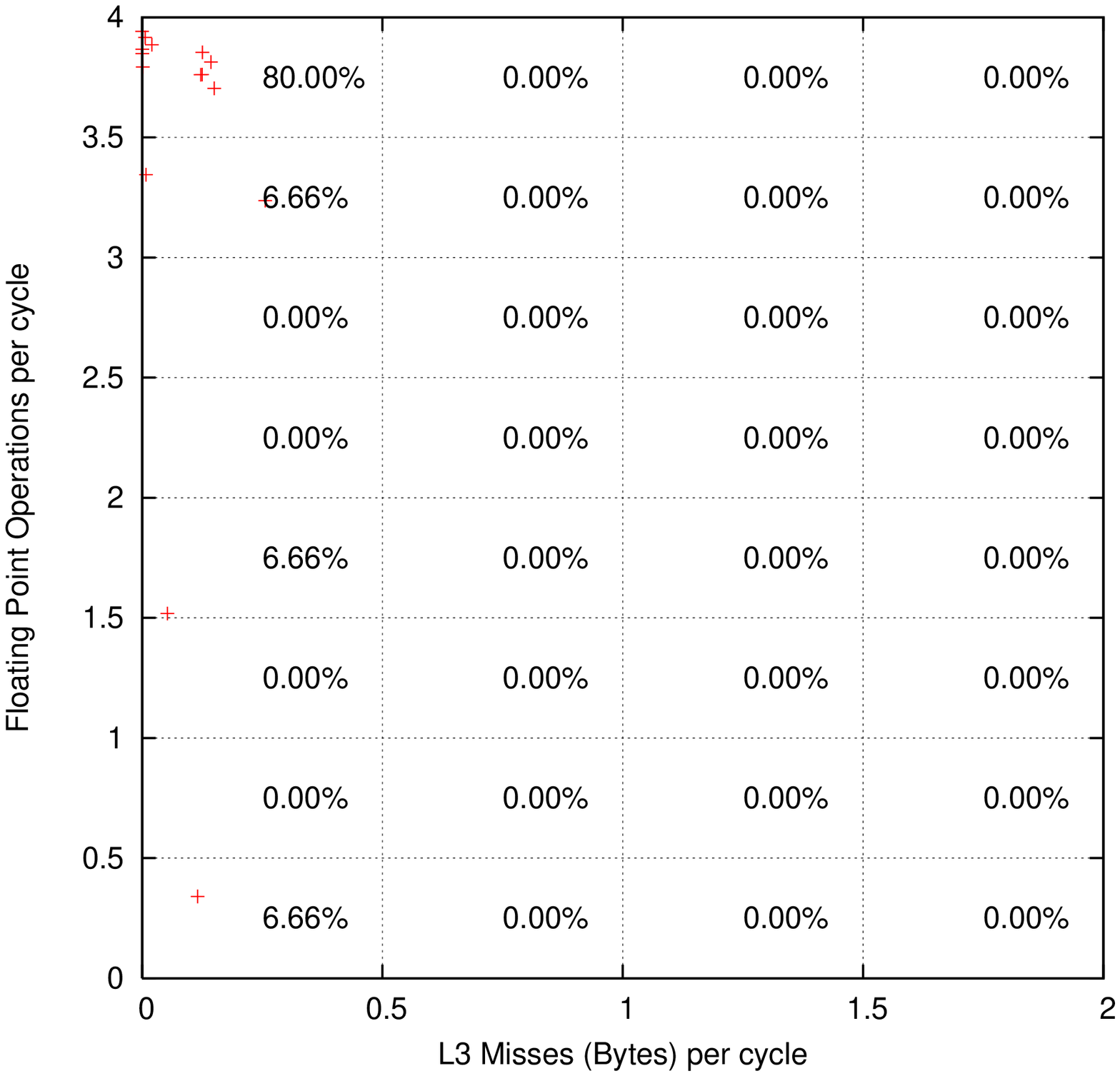}&\includegraphics[scale=.26,angle=-90]{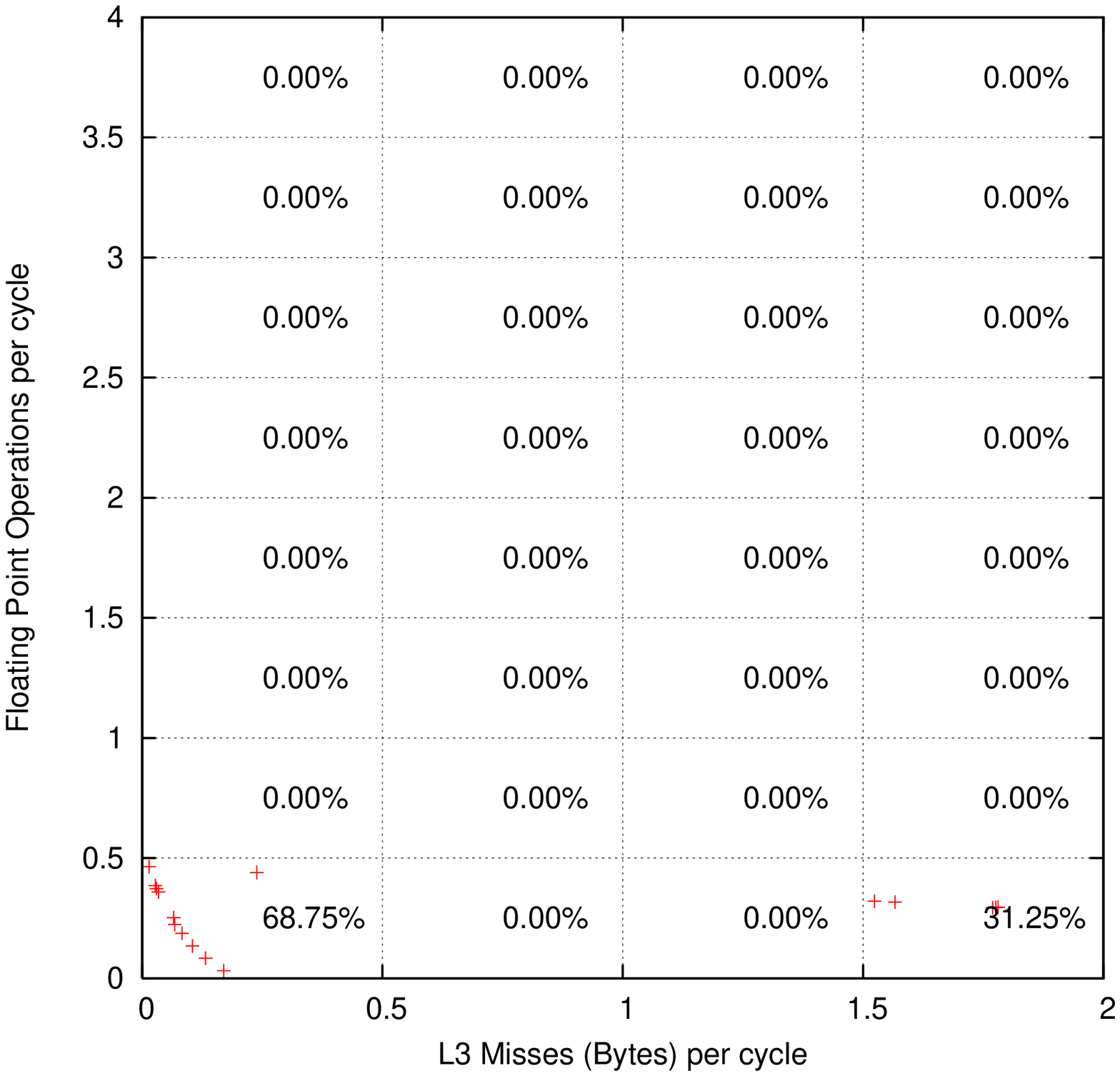} \\
(a) & (b)
\end{tabular}
\caption{Floating point operations per cycle versus L3~Misses (in Bytes) as measured by the hardware
  counters on HLRB~II for mod2am (a) and mod2as (b) (Step~2). The numbers indicate the resulting weights associated with 
  each rectangle (Step~3). 
  }
\label{fig:weightMod2}    
\end{figure}

Validating Apex-MAP by using the two mathematical kernels needs several steps:
\begin{enumerate}
\item Measure the performance of mod2am/as on the original hardware (HLRB~II).
\item Measure the hardware counters for mod2am/as on HLRB~II.
\item Generate weights for each rectangle and each kernel.
\item Measure the performance of mod2am/as on the target hardware (Nehalem EP).
\item Run Apex-MAP with the weights for mod2am/as on Nehalem and HLRB~II. 
\item Compare the predicted results (Step~5) with the actual results (Steps~1~and~4).
\end{enumerate}

Step~1 yields a mean performance on HLRB~II for all reference input data sets of 5.4~GFlops per core for mod2am (84\% of peak) 
and 0.5~GFlops for mod2as (8\% peak). 
Figure~\ref{fig:weightMod2} shows the hardware counter measurements done in Step~2 and the derived
weights for the Apex-MAP runs (Step~3): It can be clearly seen, that the dense matrix-matrix multiplication
(a) is compute bound while the sparse matrix-vector multiplication (b) is memory bound.

\begin{figure}[h!]
\begin{tabular}{cc}
\includegraphics[scale=.26,angle=-90]{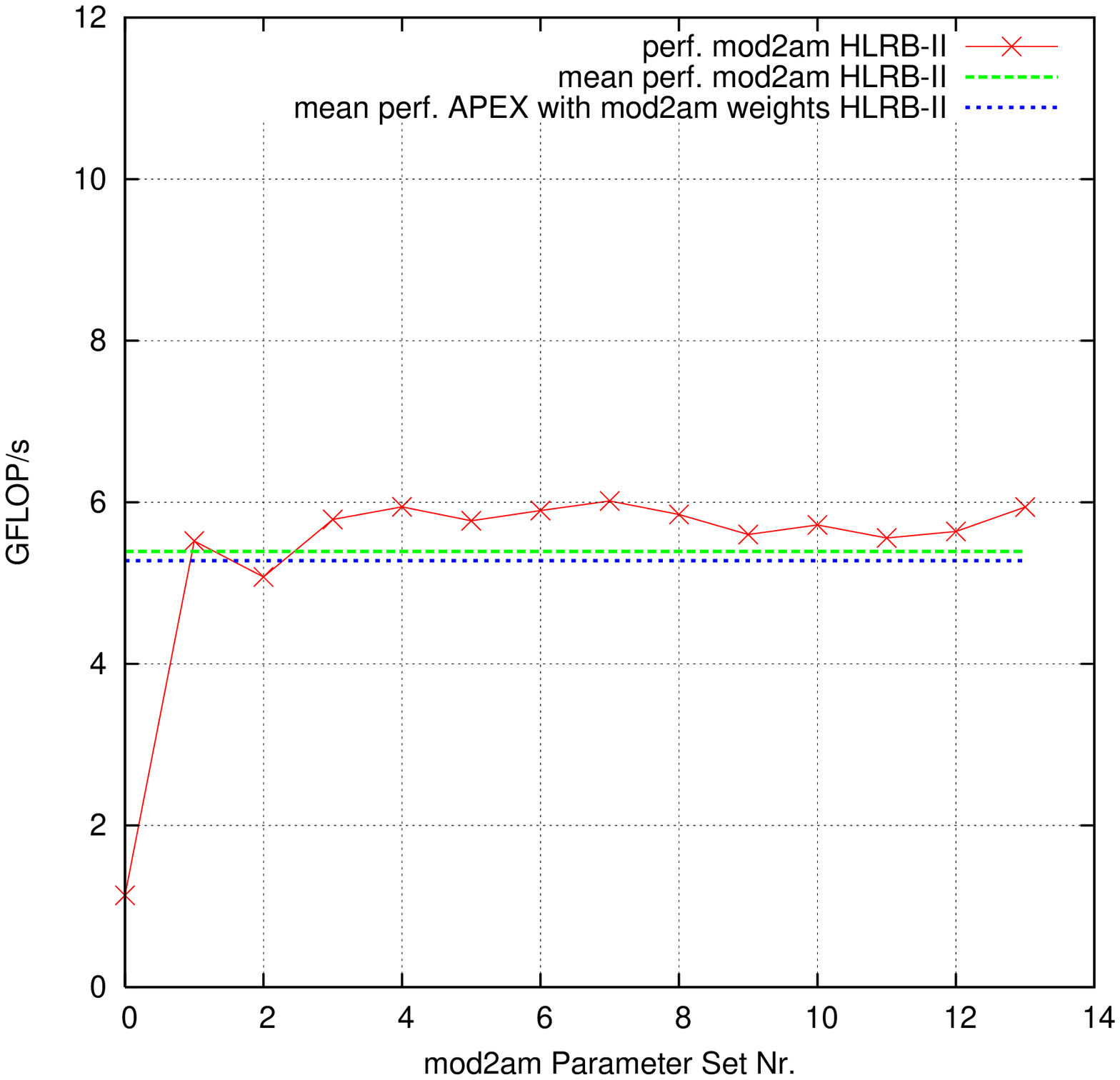}&\includegraphics[scale=.26,angle=-90]{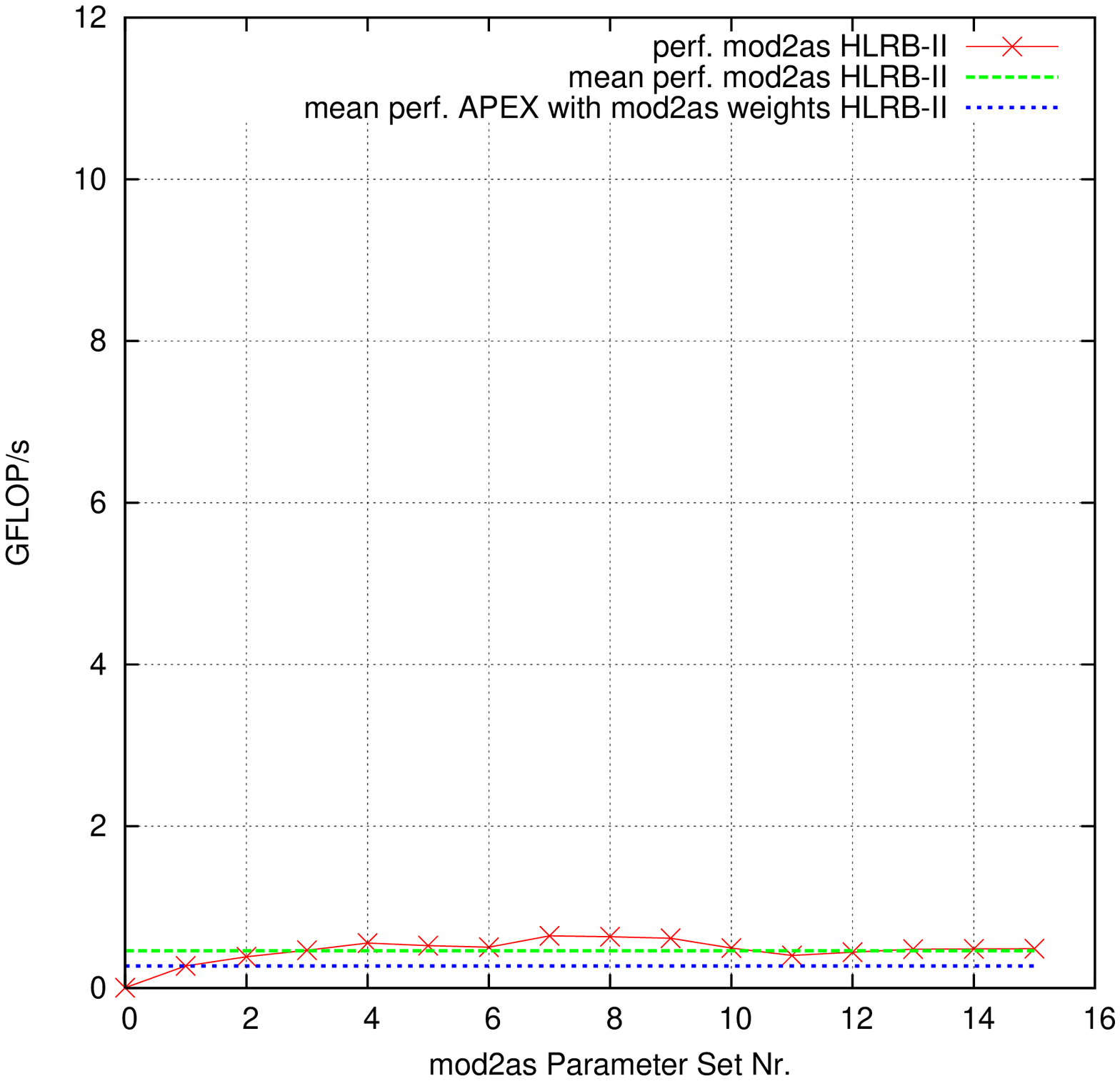} \\
(a) & (b) \\
\includegraphics[scale=.26,angle=-90]{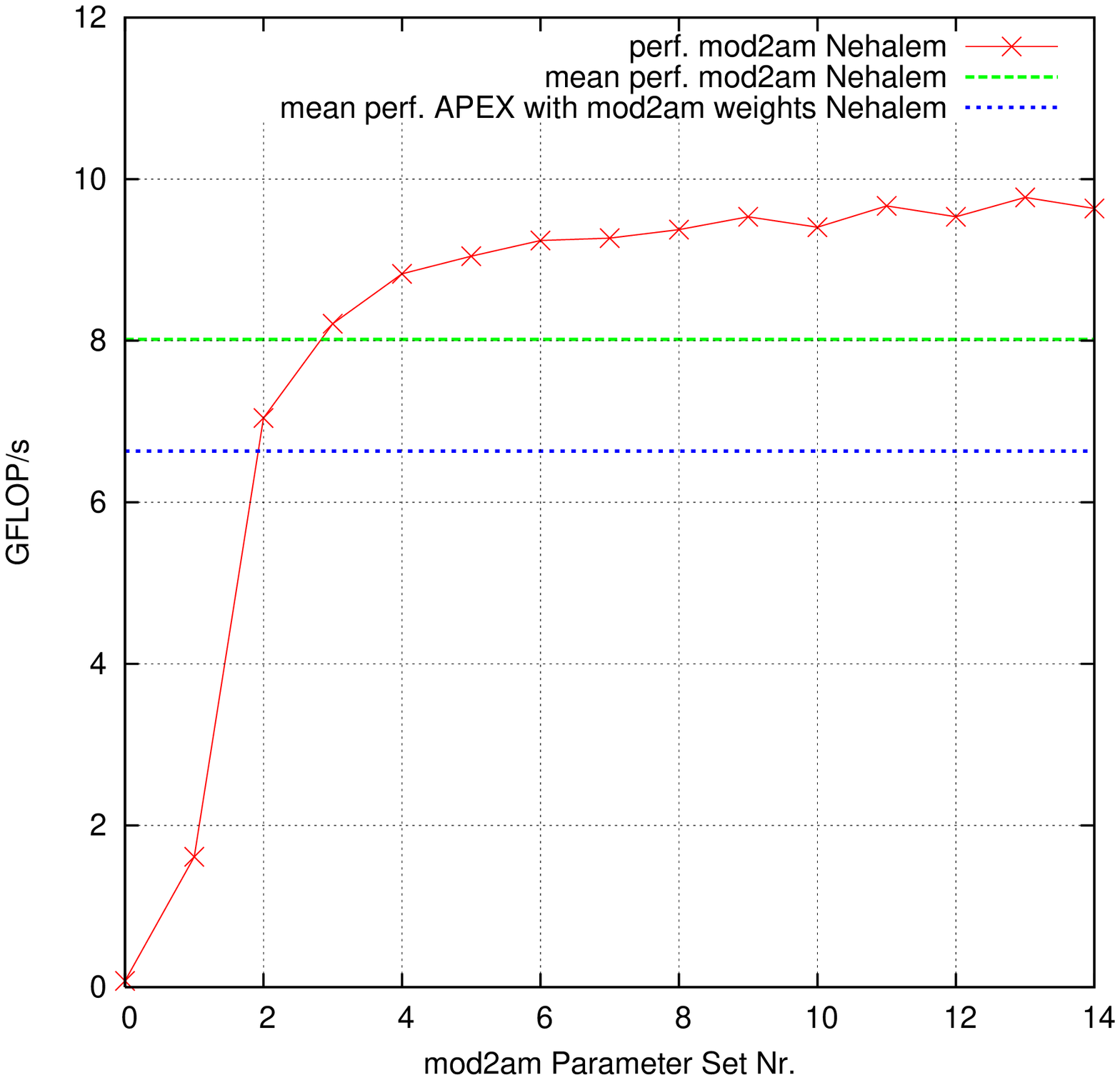}&\includegraphics[scale=.26,angle=-90]{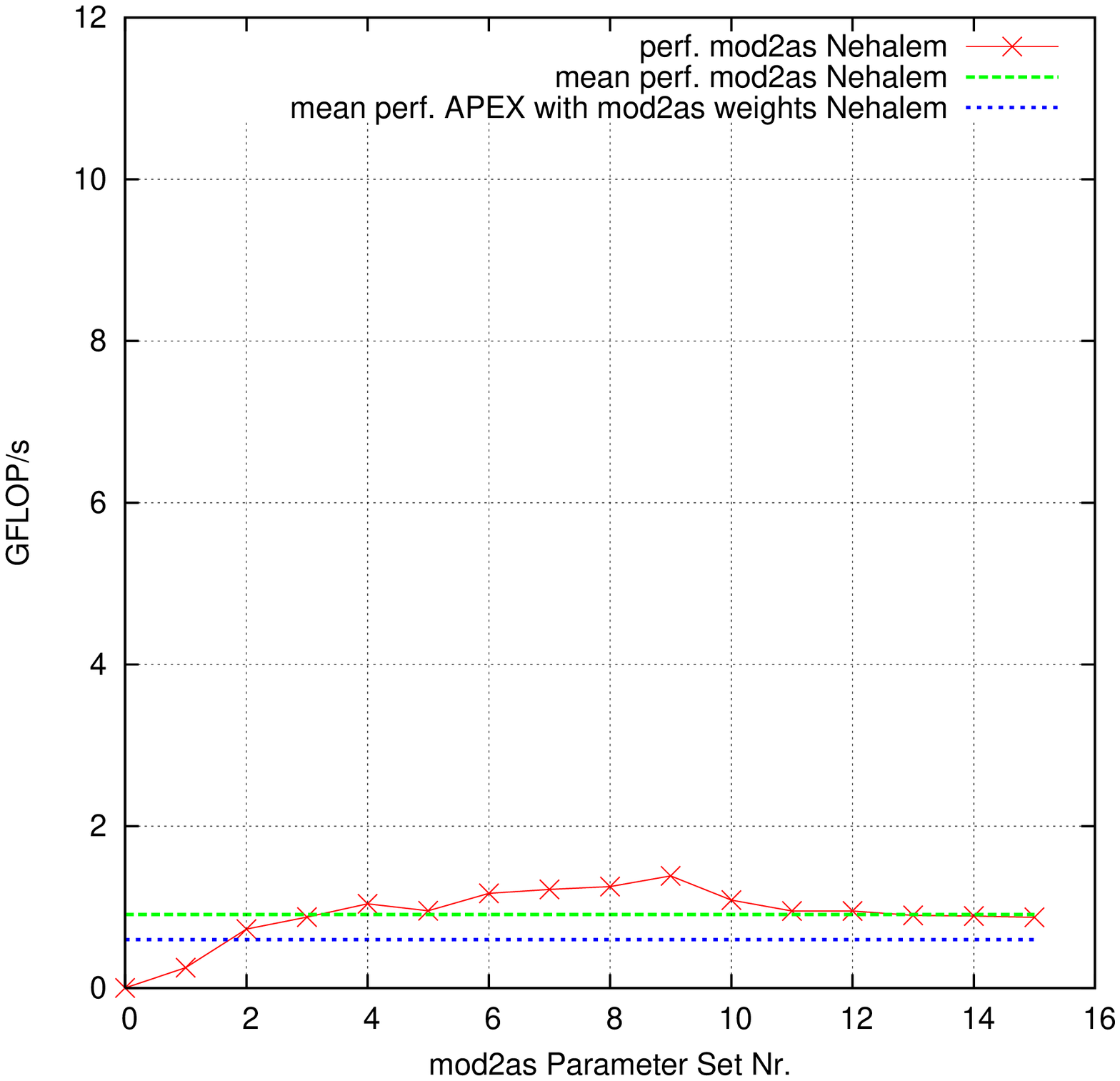} \\
(c) & (d)
\end{tabular}
\caption{All figures show the actual performance measured for each reference input data set (red curve) together with
  the mean performance measured by the mathematical kernel (green line) and predicted by Apex-MAP (blue line). 
  The first line shows results on HLRB~II (a,b), the second line results on Nehalem EP (c,d). The left diagrams
  are based on mod2am (a,c) and the diagrams on the right on mod2as (b,d).
  }
\label{fig:predMod2}  
\end{figure}

Step~4 shows an actual performance on the target architecture Nehalem EP of 8.0~GFlops (80\%) for mod2am and 0.9~GFlops (9\%) for mod2as.
The results of Step~5 can be seen in Fig.~\ref{fig:predMod2}; the mean performance predicted by Apex-MAP deviates only slightly from the
measured data on both architectures. The measurements on Nehalem are slightly worse, since the \texttt{compute} routine, which
has been optimised for Itanium is not able to reach peak performance on
Nehalem.


\section{Conclusion and Outlook}

It has been shown that an adaptation of the Apex-MAP benchmark can be used to
model the application mix in order to use it for benchmarking the suitability 
of new architectures for a computing centre. The adapted benchmark has been 
validated by using it to predict the performance of two mathematical kernels on 
two architectures.

Future work will go mainly into two directions. Firstly we want to investigate in 
more detail the quality of the predictions to refine the benchmark and ensure that
it adapts easily to new environments. Secondly we want to use Apex-MAP to investigate
if hardware accelerators are advantageous for our application mix. 
Hardware accelerators
like GPUs and the CELL processor with an enormous peak performance  
have recently gained much interest in the community. 
A programming model that was evaluated at LRZ is the multi-core development
platform RapidMind, which is a tool that allows generating code
for GPUs, the CELL processor and multi-core CPUs with the same source file. 
Using a RapidMind version of the Apex-MAP benchmark could offer an easy way to simulate typical application
performance patterns on a broad range of architectures. 
RapidMind ports of the two mathematical kernels are already available and could be used to
validate a RapidMind Apex-MAP version. 

\section*{Acknowledgment}

This work was financially supported by the KONWIHR-II project ``OMI4papps''
and by the PRACE project funded in part by the EU's 7th Framework Programme
(FP7/2007-2013) under grant agreement no.~RI-211528.

\bibliographystyle{spmpsci}

\end{document}